\documentclass[showpacs,10pt]{revtex4}
\usepackage{graphics}

\begin{document}

\title{INTERACTION OF COUPLED HIGHER ORDER NONLINEAR SCHR\"ODINGER
 EQUATION SOLITONS}

\author{Abhijit Borah}
\email{abhijit@iitg.ernet.in}
\author{Sasanka Ghosh}
\email{sasanka@iitg.ernet.in}
\author{Sudipta Nandy}
\email{sudipta@iitg.ernet.in}
\affiliation{Indian Institute of Technology, Guwahati}
\date{\today}

\begin{abstract}
The novel inelastic collision properties of two-soliton interaction
for an $n$-component coupled higher order nonlinear Schr\"odinger
 equation
are studied. Some interesting features of three soliton
 interactions, related
to the integrability of the $n$-component coupled higher order
 nonlinear
Schr\"odinger equation  are also discussed.
\end{abstract}


\maketitle

\newcommand\rf[1]{(\ref{eq:#1})}
\newcommand\lab[1]{\label{eq:#1}}
\newcommand\nonu{\nonumber}
\newcommand\br{\begin{eqnarray}}
\newcommand\er{\end{eqnarray}}
\newcommand\be{\begin{equation}}
\newcommand\ee{\end{equation}}
\renewcommand\({\left(}
\renewcommand\){\right)}
\renewcommand\v{\vert}                     
\newcommand\bv{\bigm\vert}               
\newcommand\Bgv{\;\Bigg\vert}            
\newcommand\bgv{\bigg\vert}              
\newcommand\lskip{\vskip\baselineskip\vskip-\parskip\noindent}
\newcommand\mskp{\par\vskip 0.3cm \par\noindent}
\newcommand\sskp{\par\vskip 0.15cm \par\noindent}
\newcommand\bc{\begin{center}}
\newcommand\ec{\end{center}}
\newcommand\Lbf[1]{{\Large \textbf{{#1}}}}
\newcommand\lbf[1]{{\large \textbf{{#1}}}}


\renewcommand\a{\alpha}
\renewcommand\b{\beta}
\renewcommand\c{\chi}
\renewcommand\d{\delta}
\newcommand\D{\Delta}
\newcommand\eps{\epsilon}
\newcommand\vareps{\varepsilon}
\newcommand\g{\gamma}
\newcommand\G{\Gamma}
\newcommand\grad{\nabla}
\newcommand\h{{1\over 2}}
\renewcommand\k{\kappa}
\renewcommand\l{\lambda}
\renewcommand\L{\Lambda}
\newcommand\m{\mu}
\newcommand\n{\nu}
\renewcommand\o{\over}
\newcommand\om{\omega}
\renewcommand\O{\Omega}
\newcommand\p{\phi}
\newcommand\vp{\varphi}
\renewcommand\P{\Phi}
\newcommand\pa{\partial}
\newcommand\tpa{{\tilde \partial}}
\newcommand\bpa{{\bar \partial}}
\newcommand\pr{\prime}
\newcommand\ra{\rightarrow}
\newcommand\lra{\longrightarrow}
\renewcommand\r{\rho}
\newcommand\s{\sigma}
\renewcommand\S{\Sigma}
\renewcommand\t{\tau}
\renewcommand\th{\theta}
\newcommand\bth{{\bar \theta}}
\newcommand\Th{\Theta}
\newcommand\z{\zeta}
\newcommand\ti{\tilde}
\newcommand\wti{\widetilde}
\newcommand\adot{\stackrel{.}{\a}}    
\newcommand\bdot{\stackrel{.}{\b}}
\newcommand\gdot{\stackrel{.}{\gamma}}
\renewcommand\ddot{\stackrel{.}{\d}}


\newcommand\cA{{\mathcal A}}
\newcommand\cB{{\mathcal B}}
\newcommand\cC{{\mathcal C}}
\newcommand\cD{{\mathcal D}}
\newcommand\cE{{\mathcal E}}
\newcommand\cF{{\mathcal F}}
\newcommand\cG{{\mathcal G}}
\newcommand\cH{{\mathcal H}}
\newcommand\cI{{\mathcal I}}
\newcommand\cJ{{\mathcal J}}
\newcommand\cK{{\mathcal K}}
\newcommand\cL{{\mathcal L}}
\newcommand\cM{{\mathcal M}}
\newcommand\cN{{\mathcal N}}
\newcommand\cO{{\mathcal O}}
\newcommand\cP{{\mathcal P}}
\newcommand\cQ{{\mathcal Q}}
\newcommand\cR{{\mathcal R}}
\newcommand\cS{{\mathcal S}}
\newcommand\cT{{\mathcal T}}
\newcommand\cU{{\mathcal U}}
\newcommand\cV{{\mathcal V}}
\newcommand\cX{{\mathcal X}}
\newcommand\cW{{\mathcal W}}
\newcommand\cY{{\mathcal Y}}
\newcommand\cZ{{\mathcal Z}}

\newcommand{\bi}[1]{\bibitem{#1}}
%
%
\newcommand\REP[3]{\textsl{Rep. Math. Phys.} \textbf{#1}, #3 (#2)}
\newcommand\PRL[3]{\textsl{Phys. Rev. Lett.} \textbf{#1}, #3 (#2)}
\newcommand\NPB[3]{\textsl{Nucl. Phys.} \textbf{B#1}, #3 (#2)}
\newcommand\JMP[3]{\textsl{J. Math. Phys.} \textbf{#1}, #3 (#2)}
\newcommand\JAP[3]{\textsl{J. Appl. Phys. } \textbf{#1}, #3 (#2)}
\newcommand\PRE[3]{\textsl{Phys. Rev. } \textbf{E#1}, #3 (#2)}
\newcommand\AP[3]{\textsl{Ann. Phys. } \textbf{#1}, #3 (#2)}


\section{Introduction\label{intro}}
The subject of integrable model is very fascinating largely because
of its innumerable symmetries and a special class of solutions
known as soliton solutions. Only a few systems in the nature are
known to be  integrable. The coupled nonlinear Schr\"odinger
equation (CNLS) and its higher  order generalization namely coupled
higher order nonlinear Schr\"odinger  equation (CHNLS) are some of
the examples of integrable equations which have  direct relevance
in the propagation of optical solitons in Kerr type  nonlinear
fibre. Coupled integrable systems have many important applications
in photorefractive crystals \cite{1} and also in all optical
computations \cite{2}. The cross phase modulation (CPM) phenomena
in CNLS  equation provides an interesting pulse shepherding effect
to align the time arrival  of the pulses \cite{3}. The CPM
phenomena along with group velocity dispersion  (GVD) can also be
utilized in compressing optical pulses at the one soliton level
\cite{4}. Recently the existence of multi-component solitons are
experimentally established \cite{5}.

In this paper we have considered some interesting features of CHNLS
equation, which describes the propagation of the optical pulses of
very short  width, of the order of $10^{-15 }$sec. We have studied
the novel inelastic collision of two solitons for an  $n$-component
CHNLS equation. The inelastic collisions among the three solitons
are also studied. The three soliton interactions may be interpreted
as a combination of two-soliton interactions occuring at three
different points.  This is in compliance with the three particle
interactions in integrable  systems. The energy exchange among the
components of three solitons may occur at  any of the three
two-soliton interaction points. In this context, it is  important
to note that the phenomena of shape changing vis a vis inelastic
collisions and their consequences have been observed for the CNLS
equation by  considering two soliton solutions of both two and
three components and  subsequently $n$ component generalization of
two soliton solutions and n-complexes  have been studied \cite{2}.
But the soliton solution being nonperturbative,  the shape changing
phenomena of CNLS equation does not ensure the same for  CHNLS
equation. Moreover, from integrability point of view study of the
collisions of three solitons becomes indispensable and one needs to
address  this issue seperately.

The paper is organized in the following sequence. In section
\ref{intro} the CHNLS equation is introduced and its $N$ soliton
solution is given explicitly. In section \ref{sec:3} two-soliton
interactions and its asymptotic analysis are obtained showing the
novel inelastic  collisions among the components of the solitons.
Some interesting features of  three soliton interactions, related
to the integrability of the  $n$-component CHNLS equation are
addressed in section \ref{sec:4}. Section  \ref{sec:5} is
the concluding one.

\section{CHNLS equation and N-soliton solution} \label{sec:2}
The $n$-component CHNLS equation incorporating the effects of Kerr
type nonlinearity and stimulated Raman scattering may be written as
\be
\label{A.11}
E_{iz} + iE_{i\t\t} + 2i (\sum_{j=1}^{n} E_j^*E_j) E_{i}
+\vareps E_{i\t\t\t} + 3\vareps (\sum_{j=1}^{n}
E_j^*E_j) E_{i\t} +3\vareps(\sum_{j=1}^{n} E_j^*E_{j\t}) E_i = 0
\ee
where $E_i$ is $n$-component electric field. $z$ and $\tau$ denote
the direction of propagation and time variable respectively. The
parameter  $\vareps$ is the ratio of the width of the spectra $
{\Delta \omega}$ to the carrier frequency ${\omega}$ such that
$\vareps=\frac{\D\om}{\om}<1$.

A particular gauge equivalent form of (\ref{A.11}) may be written
as \cite{6}
\be
\label{A.12}
q_{ix} + \vareps q_{ittt} + 3\vareps \sum_{j=1}^{n} (q_j^*q_j)
q_{it} + 3\vareps \sum_{j=1}^{n}(q_j^*q_{jt}) q_i = 0
\ee
where the transformation relations are given by
\br
\label{A.13}
& &E_i(z,\t)=q_i(x,t)e^{i(-\frac{1}{27\vareps^2}x-
\frac{1}{3\vareps}t)}\nonumber\\
& &x=z \nonumber\\
& & t=\t-\frac{1}{3\epsilon}z \nonumber
\er
The transformed equation (\ref{A.12}) is known as the coupled
complex modified KdV equation (CCMKdV) whose $N$ soliton solutions
for $n$-component  field is known \cite{6}. Notice that the CCMKdV
equation in (\ref{A.12}) is  convenient particularly for the Lax
representation of the system and  consequently for applying the
inverse scattering method to obtain $N$-soliton solution. However,
as a result of the transformations (\ref{A.13}), there is no change
of the envelope function,  but for a constant shift in velocity for
all solitons. Consequently, the shape of the pulses for CCMKdV
equation  and CHNLS equation remains the same \textit{i.e.,}
$|E_i|=|q_i|$ and the  transformation will not affect the
subsequent results of the two gauge equivalent  systems.

The $N$- soliton solution of the $n$ component CCMKdV equation may
be written in the most compact form as \cite{6},
\be
q_i(x,t) =
 -2\sum^{N}_{j=1}(\textbf{B}\textbf{C}^{-1})_{ij}e^{-i\l_j^*t}
\label{A.14}
\ee
where, $\textbf{B}$ and $\textbf{C}$ are respectively $n\times N$
and $N\times N$ matrices whose explicit forms are given by
\be
(\textbf{B})_{ij} = i {\cC}^{(j)}_{n+1i}(0)
e^{-8i\vareps\l_j^{*3}x-i\l_j^{*}t}
\label {A.15}
\ee
and
\be
\label{A.16}
(\textbf{C})_{ij} = \sum_{p=1}^n\sum_{k=1}^N
 \frac{{\cC}^{(j)}_{n+1p}(0)
{\cC}^{*(k)}_{n+1p}(0)}{\a^{\prime}_{n+1n+1}(\l_j^*)
\a^{*\prime}_{n+1n+1}(\l_k)}\cdot \frac{e^{-i(\l_i^*+\l_j^*-2\l_k
 )t+
8i\vareps(\l_k^3-\l_j^{*3})x}}{(\l_k-\l_j^*)(\l_k-\l_i^*)}-\d_{ij}
\ee
In the above equation $\a_{ij}$ are the elements of the
$(n+1)\times (n+1)$ scattering matrix and $\l$ is the spectral
parameter.  $\cC_{(n+1)p}^{(j)}$ is related to the elements of the
scattering matrix at the position  of the simple poles,
$\cC_{(n+1)p}^{(j)}=\alpha_{(n+1)p}(\l_j^*)|_{x=0}$ and $\prime$
over $\alpha$ denotes the derivative with respect to $\l$.

The one soliton solution (1SS) for n-coupled system directly
follows from (\ref{A.14},\ref{A.15},\ref{A.16}) by considering
$N=1$ and is  given as
\be
q_i(x,t) = \frac{{\cC}^{(1)}_{n+1i}e^{-\frac{R_{11}}{2}}
e^{i\eta_{1I}}}{\cosh(\eta_{1R}+\frac{R_{11}}{2})}
\label{A.17}
\ee
where $\eta_{1R}$ and $\eta_{1I}$ respectively denote the real and
imaginary parts of $\eta_1=-2i\l_1^*t-8i\vareps \l_1^{*3} x
+i\frac{\pi}{2}$ and $e^{R_{11}}=-\frac{\k_{11}}{\l^2_{11}}$. The
complex variables are defined as
\[{\k}_{11} = \displaystyle{\sum_{p=1}^n}
\Big|\frac{{\cC}^{(1)}_{(n+1)p}}{{\a}'_{n+1 n+1}(\l_1)}\Big|^2 \nonumber
\]
and
\[
 \l_{11}=\l_1-\l^*_1
\]
For convenience let us define $\l_1=\frac{\l_{1R} +i \l_{1I}}{2}$,
where the  subscripts $R$ and $I$ denote the real and imaginary
parts and consequently  $\eta_{1R}$ and $\eta_{1I}$ become
\be
\eta_{1I}=-\l_{1R}t + \vareps\l_{1R}(3\l_{1I}^2- \l_{1R}^2)x
+\frac{\pi}{2}
\label{A.18}
\ee
\be
\eta_{1R}=-\l_{1I}[t - \vareps(\l_{1I}^2- 3\l_{1R}^2)x]
\label{A.19}
\ee
From the argument of the $\cosh$ function of the 1SS (\ref{A.17})
and (\ref{A.19}), it is straight forward to identify the width of
the  soliton pulse as $\Gamma=|\l_{1I}|^{-1}$ and the soliton
travels with  a group velocity
$V_g=[\epsilon(\l_{1I}^2-3\l_{1R}^2)]^{-1}$ in  the positive
$x$ direction when $\l_{1I}^2>3\l_{1R}^2$ and in the  negative $x$
direction when $\l_{1I}^2<3\l_{1R}^2$.

\section{ Two-Soliton interaction}\label{sec:3}
The two soliton solution (2SS) can be obtained from
(\ref{A.14},\ref{A.15},\ref{A.16}) by putting $N=2$, the explicit
form of the $i^{ih}$ component of 2SS being
\be
q_i=\frac{2\sum^2_{m=1} \Big( \cC^{(m)}_{n+1 i}e^{\eta_m}
+e^{\eta_1+ \eta_2 +\eta_m^\star +\delta_m^{(i)}}\Big)}{Det[C]}
\label{A.26}
\ee
where,
\be
e^{\d^{(i)}_{m}}=\sum^2_{r,s=1}{\cC}^{(r)}_{n+1
i}\frac{\k_{ms}}{\l_{ms}}
\bigg(\frac{1}{\l_{mr}}-\frac{1}{\l_{ms}} \bigg)
\label{A.27}
\ee
In (\ref{A.27}) $\l_{ij}=\l_i-\l_j^*$ and
$\eta_m=-2i\l^*_mt - 8i\vareps\l^{*3}_mx+i\frac{\pi}{2}$.
$Det[C]$ is the determinant of $2\times 2$ dimensional matrix
obtained from $(\textbf{C})_{ij}$ (\ref{A.16}). The form of
$Det[C]$ in terms of the spectral parameters emerges rather lengthy
but it  will be useful for the asymptotic analysis of the
interacting solitons and will be  seen later. In terms of the
spectral parameters $Det [C]$ is given as
\be
Det[C]=1+\sum^2_{r,s=1}e^{\eta_r+\eta_s^\star+R_{rs}}
+e^{2\eta_{1R}+2\eta_{2R}+{\cQ}} \label{A.28}
\ee
with
\be
e^{R_{rs}}=-\frac{\k_{rs}}{\l^2_{rs}},
\label{A.29}
\ee
\be
\k_{rs}=\sum^{n}_{p=1}
\frac{{\cC}^{(s)}_{n+1 p}{\cC}^{\star(r)}_{n+1 p}}
{\alpha_{(n+1)(n+1)}'(\l_r^\star)\a'^\star_{(n+1)(n+1)}(\l_s)}
\label{A.30}
\ee
and
$e^{\cQ}=Det[\k]Det[\cL],[\k]=(\frac{\k_{ij}}{\l_{ij}})$ (no sum
over $i,j$), $[{\cL}]=(\l_{ij})^{-1}$, $[\k]$ and $[{\cL}]$ being
$2\times 2$ matrices. To understand the interaction in a more
explicit manner we analyse  the asymptotic limits of the 2SS
(\ref{A.26}). The asymptotic limit may be obtained by observing the
2SS when both the solitons are infinitely  apart. This may be
achieved by taking the limit $\eta_{2R}\rightarrow \pm  \infty$. As
a consequence $i^{th}$ component of the remaining soliton  acquires
the form
\be
q_i^{(l\pm)}=\frac{\l_{lI}A_i^{(l\pm)}e^{i\eta_{lI}}}
{\cosh(\eta_{lR}+\Phi^{l\pm})}
\label{A.32}
\ee
where $A_i^{(l\pm)}$ and $\Phi^{(l\pm)}$ may be interpreted as the
amplitudes and phases of $i^{th}$ component of $l^{th}$ soliton
respectively. Notice that in general $A_i^{(l-)}$ are different
from $A_i^{(l+)}$. This may occur due to exchange of energy between
the solitons. The amplitude $A_i^{(l-)}$ in terms of
 ${\cC}_{n+1i}^{(l)}$
is given as
\be
A^{(l-)}_i=\frac{{\cC}_{n+1i}^{(l)}}{\sqrt{\k_{ll}}}
\label{A.33}
\ee
The expression for $A_i^{(l+)}$ however, is more involved and may
be written in terms of $A_i^{(l-)}$ as
\be
A_i^{(l+)}=A_i^{(l-)}T^l_i
\label{A.34}
\ee
$T^l_i$ may be interpreted as the transition matrix and is defined
as
\be
T^1_i=\frac{(1-\frac{{\cC}_{n+1i}^{(2)}}{{\cC}_{n+1i}^{(1)}}\L_2)}
{\sqrt{1-\L_1\L_2}}\sqrt{(\frac{\l_2^*-\l_1^*}{\l_2-\l_1})
(\frac{\l_2^*-\l_1}{\l_2-\l_1^*})}
\label{A.35}
\ee
and
\be
T^2_i=\frac{(1-\frac{{\cC}_{n+1i}^{(1)}}{{\cC}_{n+1i}^{(2)}}\L_2)}
{\sqrt{1-\L_1\L_2}}\sqrt{(\frac{\l_2^*-\l_1^*}{\l_2-\l_1})
(\frac{\l_2 -\l_1^*}{\l_2^*-\l_1})}
\label{A.36}
\ee
where $\L_1=\frac{\k_{12}\l_{11}}{\k_{11}\l_{12}}$,
$\L_2=\frac{\k_{21}\l_{22}}{\k_{22}\l_{21}}$. If $|T^l_i|=1$, the
solitons pass through each other without being  affected in their
shapes and sizes. Otherwise, we will see that the solitons exchange
energy at the time of interaction. The phase shift $\Phi^{(l)}$ as
a result of collision may be obtained from the following relation.
\be
\Phi^{(l)}=\Phi^{(l+)}-\Phi^{(l-)}
\label{A.37}
\ee
where $\Phi^{(l-)}=\frac{1}{2}(R_{11})$ and
$\Phi^{(l+)}=\frac{1}{2}({\cQ}-R_{22})$ and consequently the phase
 shift
becomes $\Phi^{(l)}=\frac{1}{2}({\cQ}-R_{11}-R_{22})$. It is now
interesting to note that for equal values of ${\cC}_{n+1i}^{(l)}$
for each component of the solitons, $|T^l_i|$  becomes unity and
consequently there is no exchange of energy among the components of
each soliton.  However, if the relative phases among
${\cC}_{n+1 i}^{(l)}$ are  introduced the energy of each individual
component no longer remains constant due  to collisions and in that
case $|T^l_i|\neq 1$. This analysis has been  presented graphically
by plotting  two soliton interaction for a two component system
with the $\vareps=0.1$, $\l_1=-1.5-i$ and $\l_1=-.5-2i$.  One of
the solitons is moving with group velocity $-1.74$ and the other is
moving with a group velocity  $3.07$.
\begin{figure}
\begin{center}
\resizebox{0.80\textwidth}{!}{%
  \includegraphics{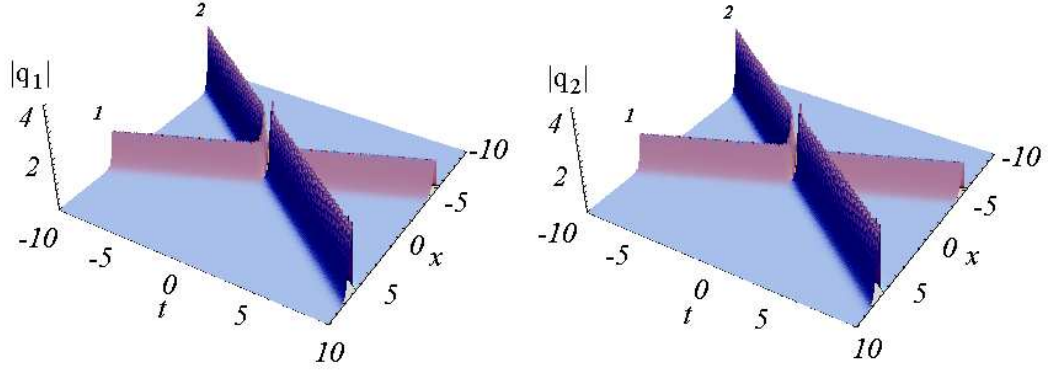}
}
\caption{Plot of $|q_1|$ and $|q_2|$ showing elastic collision}
\label{fig:1}
\end{center}
\end{figure}
\begin{figure}
\begin{center}
\resizebox{0.80\textwidth}{!}{%
  \includegraphics{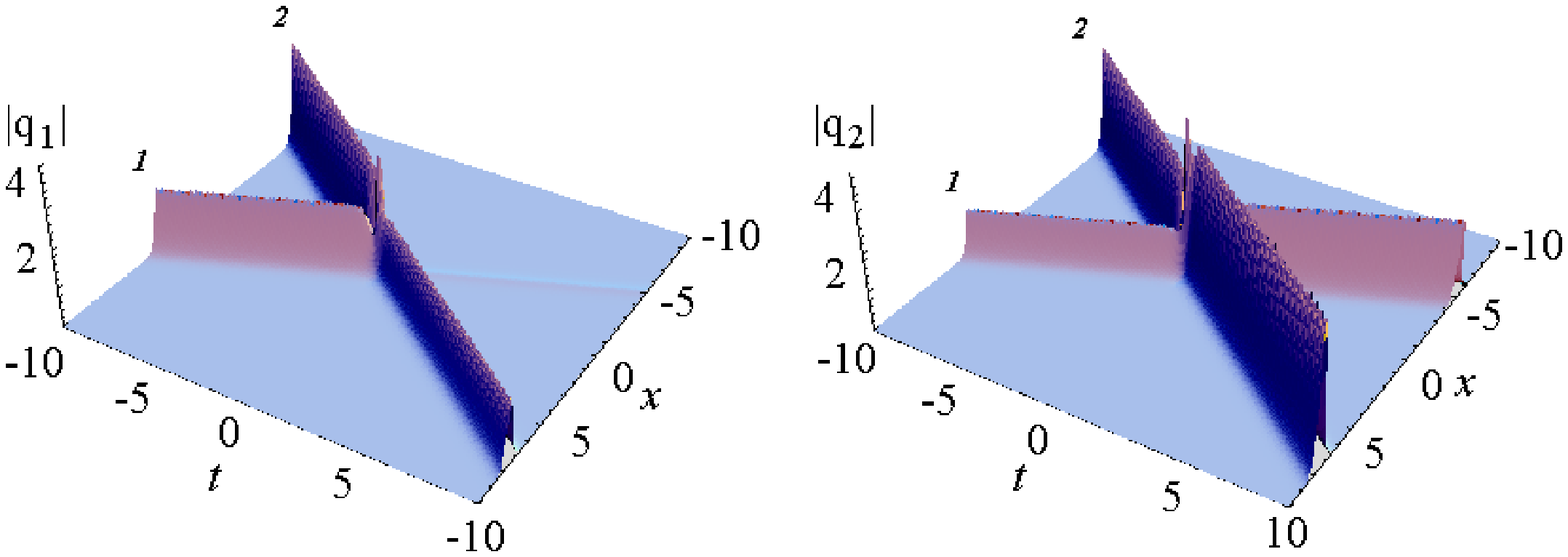}
}
\caption{Plot of $|q_1|$ and $|q_2|$ showing inelastic collision }
\label{fig:2}
\end{center}
\end{figure}
\begin{figure}
\begin{center}
\resizebox{0.40\textwidth}{!}{%
  \includegraphics{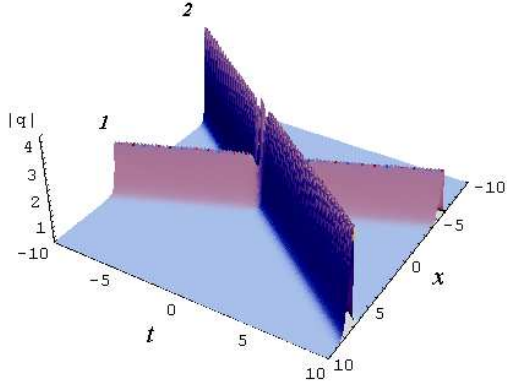}
}
\caption{Plot of $|q|$ showing constant energy profile }
\label{fig:3}
\end{center}
\end{figure}
In \textit{figures}(\ref{fig:1}) two soliton interaction is
plotted with ${\cC}_{n+1 i}^{(l)}=1$ for each component, where the
energy profile of each component of the solitons remains unchanged.
In \textit{figures}(\ref{fig:2}) two soliton interaction is plotted
with  nontrivial relative phases,
${\cC}_{n+1\ 1}^{(1)}={\cC}_{n+1\ 2}^{(1)}={\cC}_{n+1\
 2}^{(2)}=1$
and ${\cC}_{n+1\ 1}^{(2)}=46 (1-i)$ showing a considerable
exchange of energy among the  components.
 \textit{figure}(\ref{fig:3})
shows that the overall intensity profile, represented by $|q|$ of
the 2SS remains unchanged although there are  appreciable  energy
exchange among the components of a soliton. It is an interesting
result in the context of all optical computations \cite{2} leading
to the  construction of the logical binary gates.

\section{Three-soliton interactions} \label{sec:4}
\begin{figure}
\begin{center}
\resizebox{0.60\textwidth}{!}{%
  \includegraphics{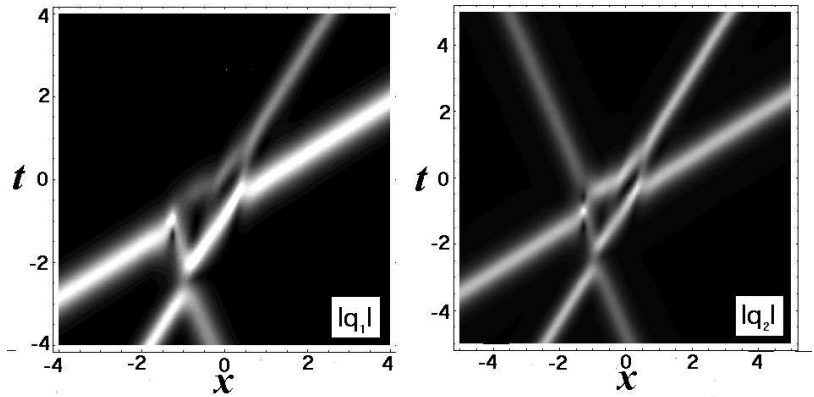}
}
\caption{Contour plot of three soliton interaction}
\label{fig:4}
\end{center}
\end{figure}

The explicit form of the  three soliton solution (3SS) can be
obtained from  (\ref{A.14},\ref{A.15},\ref{A.16}) by putting $N=3$,
the explicit form of the $i^{ih}$ component of 3SS being
\begin{eqnarray*}
q_p=\frac{-2}{Det[C]}\Bigg[\sum_{i=1}^3{\cC}_{n+1
p}^{(i)}e^{\eta_i}+\sum_{i,j,m=1_{i\ne j\ne m}}^3 (adj
{\tilde \k}^j)_{mm} (adj {\tilde {\cL}}^j)_{mm}
e^{2\eta_{iR}+\eta_j}+\sum_{m=1}^3Det[{\tilde \k}^m]
Det[{\tilde {\cL}}^m]
e^{\displaystyle{\sum_{i,m=1\atop i\ne m}^3}
{2\eta_{iR}+\eta_m}}\Big] \label{4.1}
\end{eqnarray*}
where,
\begin{eqnarray*}
Det[C]=-1-\sum_{i,j=1}^3e^{\eta_i + \eta_j^*}e^{R_{ij}}
-\sum_{i=1}^3\sum_{k,j=1\atop i\ne j \ne k}^3(adj\k)_{jk}
(adj{\cL})_{jk}e^{2\eta_{iR}+\eta_k+\eta_j^{\star}}
- \sum_{i,j,k=1\atop i\ne j\ne k}^3(adj\k)_{kk}(adj{\cL})_{kk}
e^{2\eta_{iR}+2\eta_{jR}}\nonu \\
+ Det[\k]Det[{\cL}]
e^{2\eta_{1_R}+2\eta_{2_R}+2\eta_{3_R}}
\label{4.2}
\end{eqnarray*}
with
${\cL}$ is a $3 \times 3$ dimensional matrix with elements
$(\l_{ij}^{-1})$ and $[\k]$ is also a $3\times3$
dimensional matrix with elements $\frac{\k_{ij}}{\l_{ij}}$ (no sum
over  $i,j$). [$\tilde{\k}^j$] is the matrix [$\k$], where the
$j$th row has been  replaced by ${{3\atop \sum}\atop{i=1}}e_{ji}
\cC_{n+1 p}^{(i)}$ and ${\tilde{\cL}}^j$ is the matrix where the
$j$th row has been replaced by a unit row  vector,
${{3\atop\sum}\atop i=1}e_{ji}$. The phenomena of shape changing
due to exchange of energy among the components of solitons becomes
more  interesting in the three-soliton interaction case. It is
observed that the  interactions take place at three different space
time points with two solitons  interact at each time. Thus the
energy exchage among the components in three  solitons interaction
may occur at three different points. This is  demonstrated in
\textit{ figure}(\ref{fig:4}) by the contour plot of the three
solitons interaction with $\epsilon=0.1, \l_1
 =-1.5-i,\l_2=-0.01+2i,
\l_3=-0.9+2i, {\cC}_{n+1\ 1}^{(1)}= {\cC}_{n+1\ 1}^{(2)}=
{\cC}_{n+1\ 1}^{(3)}=10^{-3}, {\cC}_{n+1\ 2}^{(2)}
={\cC}_{n+1\ 2}^{(3)}=1$ and $  {\cC}_{n+1\ 2}^{(1)}= 10^3-0.8i$.
Two of the solitons are moving in the positive $x$ direction with
group  velocities $2.5$ and $6.36$ and the third soliton is moving
in the negative  $x$ direction with group velocity $-1.74$. The
nature of three soliton  interactions opens up another possibility
of three solitons interaction where second and  third solitons
interact first instead of first and second solitons  leaving the
final configuration same in both the cases.

\section{Conclusion}\label{sec:5}
In conclusion, we have shown  the novel shape changing phenomena
asscoiated with two soliton interaction  for $n$ component CHNLS
equation. The three solitons interaction also exhibits energy
exchange, but the exchange may occur at three points. This may lead
to a more  flexibility in constructing logical gates in all optical
computing systems. The  three solitons interaction also
demonstartes an interesting feature  conforming the exact
integrability of the system \textit{ a la} Zamolodchikov  \cite{7}
and indicates the existence of Yang Baxter like relation, which
will be  publishedelsewhere.

\textbf{Acknowledgments.}
S. G. and S. N. would like to thank CSIR Govt. of India
for financial support under the project 03(0896)/99/EMR II. S. G.
 also
gratefully acknowledge the hospitality of the Organizing Committee
 of the
International Conference {\em ``Geometry, Integrability and
 Nonlinearity in
Condensed Matter and Soft Condensed Matter Physics''}, where
the present results were first reported.

\end{document}